\documentclass[sigconf]{acmart}

\usepackage{multirow}

\AtBeginDocument{
  }

\setcopyright{acmlicensed}
\copyrightyear{2026}
\acmYear{2026}
\acmDOI{XXXXXXX.XXXXXXX}
\acmConference[Conference acronym 'XX]{Make sure to enter the correct
  conference title from your rights confirmation email}{June 03--05,
  2018}{Woodstock, NY}
\acmISBN{978-1-4503-XXXX-X/2018/06}

\begin{document}

\title{RQ-GMM: Residual Quantized Gaussian Mixture Model for Multimodal Semantic Discretization in CTR Prediction}

\author{Ziye Tong}
\authornote{Equal contribution.}
\affiliation{
  \institution{Tencent}
  \city{Beijing}
  \country{China}
}
\email{ziyetong@tencent.com}

\author{Jiahao Liu}
\authornotemark[1]
\affiliation{
  \institution{Fudan University}
  \city{Shanghai}
  \country{China}
}
\email{jiahaoliu23@m.fudan.edu.cn}

\author{Weimin Zhang}
\authornote{Corresponding author.}
\affiliation{
  \institution{Tencent}
  \city{Beijing}
  \country{China}
}
\email{weiminzhang@tencent.com}

\author{Hongji Ruan}
\affiliation{
  \institution{Beijing Jiaotong University}
  \city{Beijing}
  \country{China}
}
\email{23125251@bjtu.edu.cn}

\author{Derick Tang}
\authornotemark[2]
\affiliation{
  \institution{Tencent}
  \city{Beijing}
  \country{China}
}
\email{dericck@tencent.com}

\author{Zhanpeng Zeng}
\affiliation{
  \institution{Tencent}
  \city{Guangzhou}
  \country{China}
}
\email{marcuszeng@tencent.com}

\author{Qinsong Zeng}
\affiliation{
  \institution{Tencent}
  \city{Guangzhou}
  \country{China}
}
\email{qinzzeng@tencent.com}

\author{Peng Zhang}
\affiliation{
  \institution{Fudan University}
  \city{Shanghai}
  \country{China}
}
\email{zhangpeng\_@fudan.edu.cn}

\author{Tun Lu}
\affiliation{
  \institution{Fudan University}
  \city{Shanghai}
  \country{China}
}
\email{lutun@fudan.edu.cn}

\author{Ning Gu}
\affiliation{
  \institution{Fudan University}
  \city{Shanghai}
  \country{China}
}
\email{ninggu@fudan.edu.cn}

\renewcommand{\shortauthors}{Ziye Tong, Jiahao Liu et al.}

\begin{abstract}
Multimodal content is crucial for click-through rate (CTR) prediction. However, directly incorporating continuous embeddings from pre-trained models into CTR models yields suboptimal results due to misaligned optimization objectives and convergence speed inconsistency during joint training. Discretizing embeddings into semantic IDs before feeding them into CTR models offers a more effective solution, yet existing methods suffer from limited codebook utilization, reconstruction accuracy, and semantic discriminability. We propose RQ-GMM (Residual Quantized Gaussian Mixture Model), which introduces probabilistic modeling to better capture the statistical structure of multimodal embedding spaces. Through Gaussian Mixture Models combined with residual quantization, RQ-GMM achieves superior codebook utilization and reconstruction accuracy. Experiments on public datasets and online A/B tests on a large-scale short-video platform serving hundreds of millions of users demonstrate substantial improvements: RQ-GMM yields a 1.502\% gain in Advertiser Value over strong baselines. The method has been fully deployed, serving daily recommendations for hundreds of millions of users.
\end{abstract}

\ccsdesc[500]{Information systems~Retrieval models and ranking}
\ccsdesc[300]{Computing methodologies~Machine learning}

\keywords{click-through rate prediction, multimodal recommendation, residual quantization, Gaussian mixture models}

\maketitle

\section{Introduction}
Click-through rate (CTR) prediction is central to modern recommender systems~\cite{han2025fedcia,liu2023personalized,liu2023autoseqrec}, directly impacting user experience and platform revenue~\cite{liu2025agentcf++,liu2025filtering,liu2022parameter}. As internet content ecosystems expand rapidly, multimodal information (e.g., images, text, audio) has become essential for characterizing items~\cite{liu2023triple,liu2023recommendation,liu2025unbiased}. In short-video and live-streaming platforms where massive new content emerges daily, multimodal content provides rich semantic-level representations that significantly enhance recommendation performance~\cite{liu2026distribution,liu2025improving}. Two paradigms exist for leveraging multimodal information in CTR prediction. The first constructs item indices directly from multimodal features for semantic matching in retrieval~\cite{deng2025onerec,rajput2023recommender,zhang2025evalagent,han2026feature}. However, this approach introduces high engineering complexity and limited gains when extended to ranking—the primary CTR prediction stage. The second paradigm incorporates multimodal features as additional inputs across the recommendation pipeline~\cite{sheng2024enhancing,gu2025llm,wu2025bidirectional}, integrating more naturally with existing architectures and demonstrating superior practical feasibility.


Despite this, directly feeding pre-trained modal embeddings into CTR models yields suboptimal results for two fundamental reasons. First, pre-trained models (e.g., CLIP~\cite{radford2021learning}, BERT~\cite{devlin2019bert}) optimize for universal semantic representations rather than user preference modeling, producing embeddings with substantial noise irrelevant to recommendations. Second, joint training suffers from a training dynamics mismatch: sparse categorical features (e.g., user/item IDs) exhibit the ``one-epoch phenomenon''~\cite{zhang2022towards} with rapid convergence due to high-frequency exposure, while extracting preference signals from multimodal embeddings requires longer cycles and careful denoising. This convergence disparity degrades overall performance. A more effective solution is two-stage optimization~\cite{singh2024better,luo2025qarm}: discretizing continuous embeddings into semantic IDs, then feeding these as categorical features into CTR models. This decouples multimodal semantic extraction from preference modeling. In stage one, discretization performs unsupervised compression and denoising via clustering, mapping high-dimensional vectors to a discrete codebook. In stage two, the CTR model learns interactions between semantic IDs and other categorical features, alleviating the dynamics mismatch. Moreover, semantic IDs serve as coarse-grained anchors: when jointly provided with continuous embeddings, they supply stable cluster-level signals that guide fine-grained learning in a coarse-to-fine paradigm.

However, existing discretization methods face critical limitations that motivate our work. VQ-VAE~\cite{van2017neural}, which combines variational autoencoders with vector quantization, frequently suffers from codebook collapse where many codebook vectors remain underutilized, limiting expressiveness in high-dimensional semantic spaces. RQ-VAE~\cite{zeghidour2021soundstream} introduces multi-level residual quantization to partially alleviate utilization issues, but requires careful balancing across levels and its hard assignment via Euclidean distance neglects distributional properties, yielding suboptimal reconstruction especially for boundary samples. RQ-KMeans takes a non-parametric approach by directly applying K-means~\cite{mcqueen1967some} to residuals, which is computationally efficient but assumes spherical clusters with equal variances—an assumption that fails for complex multimodal distributions. Its hard assignment disrupts semantic continuity for boundary samples and cannot model uncertainty, hampering fine-grained semantic capture. We propose RQ-GMM (Residual Quantized Gaussian Mixture Model), introducing probabilistic modeling to better capture the multimodal embedding space's statistical structure. By modeling a Gaussian distribution per codebook vector, RQ-GMM achieves high utilization and superior reconstruction accuracy. Its soft assignment mechanism handles boundary samples more reasonably than hard clustering, while residual quantization ensures fine-grained semantic modeling.

Comprehensive experiments on public datasets demonstrate RQ-GMM's superiority in reconstruction error, codebook utilization, and CTR performance. Critically, online A/B tests on a short-video platform serving hundreds of millions of daily users show RQ-GMM improves Advertiser Value by 1.502\% over RQ-VAE—a substantial gain in large-scale systems. RQ-GMM has been fully deployed, serving daily recommendations for hundreds of millions of users. The paper proceeds as follows: Section 2 presents preliminaries; Section 3 details RQ-GMM; Section 4 analyzes experimental results; Section 5 concludes with future directions.

\section{Preliminaries}

\subsection{VQ-VAE}

Vector Quantized Variational AutoEncoder (VQ-VAE)~\cite{van2017neural} discretizes continuous representations via encoder-decoder architecture. Given $\mathbf{x} \in \mathbb{R}^D$, the encoder produces $\mathbf{z}_e = E(\mathbf{x}) \in \mathbb{R}^d$, which is quantized by nearest codebook lookup:

\begin{equation}
\mathbf{z}_q = q(\mathbf{z}_e) = \arg\min_{\mathbf{c}_k \in \mathcal{C}} \|\mathbf{z}_e - \mathbf{c}_k\|_2^2
\end{equation}
where $\mathcal{C} = \{\mathbf{c}_1, \ldots, \mathbf{c}_K\} \subset \mathbb{R}^d$ is the codebook. The decoder reconstructs $\hat{\mathbf{x}} = D(\mathbf{z}_q)$. Training optimizes:

\begin{equation}
\mathcal{L}_{\text{VQ-VAE}} = \|\mathbf{x} - \hat{\mathbf{x}}\|_2^2 + \|\text{sg}[\mathbf{z}_e] - \mathbf{z}_q\|_2^2 + \beta \|\mathbf{z}_e - \text{sg}[\mathbf{z}_q]\|_2^2
\end{equation}
with reconstruction, codebook, and commitment losses respectively, where $\text{sg}[\cdot]$ denotes stop-gradient.

\subsection{RQ-VAE}

Residual Quantized VAE (RQ-VAE)~\cite{zeghidour2021soundstream} employs multi-level cascaded residual quantization. Given $\mathbf{z}_e = E(\mathbf{x})$, it performs $L$ levels:

\begin{equation}
\mathbf{z}_q^{(l)} = q^{(l)}(\mathbf{r}^{(l-1)}), \quad \mathbf{r}^{(l)} = \mathbf{r}^{(l-1)} - \mathbf{z}_q^{(l)}
\end{equation}
where $\mathbf{r}^{(0)} = \mathbf{z}_e$ and each level uses codebook $\mathcal{C}^{(l)} = \{\mathbf{c}_1^{(l)}, \ldots, \mathbf{c}_K^{(l)}\}$. The final representation is $\mathbf{z}_q = \sum_{l=1}^{L} \mathbf{z}_q^{(l)}$, with reconstruction $\hat{\mathbf{x}} = D(\mathbf{z}_q)$. Training minimizes:

\begin{equation}
\mathcal{L}_{\text{RQ-VAE}} = \|\mathbf{x} - \hat{\mathbf{x}}\|_2^2 + \sum_{l=1}^{L} \left( \|\text{sg}[\mathbf{r}^{(l-1)}] - \mathbf{z}_q^{(l)}\|_2^2 + \beta \|\mathbf{r}^{(l-1)} - \text{sg}[\mathbf{z}_q^{(l)}]\|_2^2 \right)
\end{equation}

\subsection{RQ-KMeans}

Residual Quantized K-Means (RQ-KMeans)~\cite{deng2025onerec} implements non-parametric residual quantization directly on $\mathbf{x} \in \mathbb{R}^D$ without encoder-decoder. With $\mathbf{r}^{(0)} = \mathbf{x}$, K-means clustering~\cite{mcqueen1967some} is applied at each level:

\begin{equation}
\mathbf{z}_q^{(l)} = \arg\min_{\mathbf{c}_k^{(l)} \in \mathcal{C}^{(l)}} \|\mathbf{r}^{(l-1)} - \mathbf{c}_k^{(l)}\|_2^2
\end{equation}

Codebook updates follow standard K-means:

\begin{equation}
\mathbf{c}_k^{(l)} = \frac{1}{|\mathcal{N}_k^{(l)}|} \sum_{\mathbf{r}^{(l-1)} \in \mathcal{N}_k^{(l)}} \mathbf{r}^{(l-1)}
\end{equation}
where $\mathcal{N}_k^{(l)}$ is the set of samples assigned to cluster $k$ at level $l$.

\section{Methods}

\subsection{RQ-GMM}

We propose Residual Quantized Gaussian Mixture Model (RQ-GMM), replacing hard clustering with GMM to better capture the statistical structure of multimodal embedding spaces via probabilistic modeling.

\subsubsection{Gaussian Mixture Quantization}

Given $\mathbf{x} \in \mathbb{R}^D$, we initialize $\mathbf{r}^{(0)} = \mathbf{x}$. At level $l$, we model $\mathbf{r}^{(l-1)}$ as a mixture of $K$ Gaussians:

\begin{equation}
p(\mathbf{r}^{(l-1)}) = \sum_{k=1}^{K} \pi_k^{(l)} \mathcal{N}(\mathbf{r}^{(l-1)} \mid \boldsymbol{\mu}_k^{(l)}, \boldsymbol{\Sigma}_k^{(l)})
\end{equation}
where $\pi_k^{(l)}$ is the mixing coefficient with $\sum_{k=1}^{K} \pi_k^{(l)} = 1$, $\boldsymbol{\mu}_k^{(l)} \in \mathbb{R}^D$ is the mean, and $\boldsymbol{\Sigma}_k^{(l)} \in \mathbb{R}^{D \times D}$ is the covariance. We adopt diagonal covariance: $\boldsymbol{\Sigma}_k^{(l)} = \text{diag}(\sigma_{k,1}^{(l)2}, \ldots, \sigma_{k,D}^{(l)2})$ for efficiency.

\subsubsection{Soft Assignment and Quantization}

GMM performs soft assignment via posterior probabilities:

\begin{equation}
\gamma_k^{(l)} = p(k \mid \mathbf{r}^{(l-1)}) = \frac{\pi_k^{(l)} \mathcal{N}(\mathbf{r}^{(l-1)} \mid \boldsymbol{\mu}_k^{(l)}, \boldsymbol{\Sigma}_k^{(l)})}{\sum_{j=1}^{K} \pi_j^{(l)} \mathcal{N}(\mathbf{r}^{(l-1)} \mid \boldsymbol{\mu}_j^{(l)}, \boldsymbol{\Sigma}_j^{(l)})}
\end{equation}

During inference, we select the component with highest posterior:

\begin{equation}
k^* = \arg\max_{k \in \{1,\ldots,K\}} \gamma_k^{(l)}, \quad \mathbf{z}_q^{(l)} = \boldsymbol{\mu}_{k^*}^{(l)}
\end{equation}

The residual updates as $\mathbf{r}^{(l)} = \mathbf{r}^{(l-1)} - \mathbf{z}_q^{(l)}$. After $L$ levels, the final representation is $\mathbf{z}_q = \sum_{l=1}^{L} \mathbf{z}_q^{(l)}$.

\subsubsection{Parameter Learning}

Parameters are learned via EM algorithm~\cite{dempster1977maximum} on $\{\mathbf{x}_i\}_{i=1}^{N}$ with $\mathbf{r}_i^{(0)} = \mathbf{x}_i$, optimizing each level sequentially. While the E-step uses soft posteriors, residual propagation uses hard assignment ($\mathbf{r}_i^{(l)} = \mathbf{r}_i^{(l-1)} - \boldsymbol{\mu}_{k_i^*}^{(l)}$ where $k_i^* = \arg\max_k \gamma_{i,k}^{(l)}$) for consistency with inference.

\textbf{E-step:} Compute posteriors:
\begin{equation}
\gamma_{i,k}^{(l)} = \frac{\pi_k^{(l)} \mathcal{N}(\mathbf{r}_{i}^{(l-1)} \mid \boldsymbol{\mu}_k^{(l)}, \boldsymbol{\Sigma}_k^{(l)})}{\sum_{j=1}^{K} \pi_j^{(l)} \mathcal{N}(\mathbf{r}_{i}^{(l-1)} \mid \boldsymbol{\mu}_j^{(l)}, \boldsymbol{\Sigma}_j^{(l)})}
\end{equation}

\textbf{M-step:} Update parameters:
\begin{equation}
\begin{aligned}
\pi_k^{(l)} &= \frac{1}{N} \sum_{i=1}^{N} \gamma_{i,k}^{(l)}, \quad
\boldsymbol{\mu}_k^{(l)} = \frac{\sum_{i=1}^{N} \gamma_{i,k}^{(l)} \mathbf{r}_i^{(l-1)}}{\sum_{i=1}^{N} \gamma_{i,k}^{(l)}} \\
\sigma_{k,j}^{(l)2} &= \frac{\sum_{i=1}^{N} \gamma_{i,k}^{(l)} (r_{i,j}^{(l-1)} - \mu_{k,j}^{(l)})^2}{\sum_{i=1}^{N} \gamma_{i,k}^{(l)}}
\end{aligned}
\end{equation}

Reconstruction quality is measured as $\text{RMSE} = \sqrt{\frac{1}{N} \sum_{i=1}^{N} \|\mathbf{x}_i - \mathbf{z}_{q,i}\|_2^2}$ where $\mathbf{z}_{q,i} = \sum_{l=1}^{L} \mathbf{z}_{q,i}^{(l)}$. Since RQ-GMM operates directly on original embeddings without encoder-decoder, EM is the sole optimization procedure.

\subsubsection{Semantic ID Generation}

After $L$ levels, each sample obtains a discrete code sequence:
\begin{equation}
\text{ID}(\mathbf{x}) = [k_1^*, k_2^*, \ldots, k_L^*]
\end{equation}
where $k_l^* \in \{1, \ldots, K\}$ indexes the selected Gaussian at level $l$. This serves as the multimodal semantic ID for CTR models.

\textbf{Advantages.} Compared with hard clustering: (1) covariance matrices characterize cluster shape and uncertainty, adapting to complex distributions; (2) soft assignment yields smoother boundary quantization, reducing discontinuities; (3) mixing coefficients $\pi_k^{(l)}$ automatically reflect data density, avoiding forced uniform assignment; (4) probabilistic modeling enables finer semantic discrimination.

\subsection{Complexity Analysis}

\textbf{Training.} For $N$ samples, one EM iteration at one level costs $O(NKD)$. With $T$ iterations and $L$ levels, total complexity is $O(TLNKD)$.

\textbf{Inference.} Per sample, computing posteriors across $L$ levels with $K$ components requires $O(LKD)$, comparable to RQ-KMeans. Diagonal covariance limits overhead.

\textbf{Space.} Each level stores $K$ means ($KD$), $K$ diagonal covariances ($KD$), and $K$ mixing coefficients ($K$), totaling $O(LKD)$, same order as RQ-KMeans.

Compared with RQ-VAE, RQ-GMM avoids encoder-decoder networks and complex gradient propagation, yielding more stable training. Though RQ-GMM has additional covariance computation per iteration versus RQ-KMeans, it converges faster with fewer iterations, resulting in comparable training time.

\subsection{Downstream CTR Application}

Semantic IDs from RQ-GMM integrate seamlessly into standard CTR models (e.g., DeepFM~\cite{guo2017deepfm}, DCN~\cite{wang2017deep}, DIN~\cite{zhou2018deep}). For item $i$ with embedding $\mathbf{x}_i \in \mathbb{R}^D$, RQ-GMM generates $\text{ID}_i = [k_1^*, \ldots, k_L^*]$. The original embedding is transformed via MLP: $\mathbf{e}_{\text{modal}} = \text{MLP}(\mathbf{x}_i) \in \mathbb{R}^{d_e}$. Features are constructed as:
\begin{equation}
\mathbf{f}_i = [\text{user\_id},\; \text{item\_id},\; \text{context},\; k_1^*,\; \ldots,\; k_L^*,\; \mathbf{e}_{\text{modal}}]
\end{equation}

This coarse-to-fine design provides stable categorical anchors (semantic IDs) that guide learning from continuous embeddings, addressing the training dynamics mismatch discussed earlier. Each ID $k_l^*$ is embedded as $\mathbf{e}_l = \mathbf{W}_l[k_l^*] \in \mathbb{R}^{d_e}$ with learnable $\mathbf{W}_l \in \mathbb{R}^{K \times d_e}$. Multi-level embeddings are fused: $\mathbf{e}_{\text{semantic}} = \text{Concat}(\mathbf{e}_1, \ldots, \mathbf{e}_L)$ or $\mathbf{e}_{\text{semantic}} = \sum_{l=1}^{L} \mathbf{e}_l$.

The CTR model predicts $\hat{y} = \text{CTR\_Model}([\mathbf{e}_{\text{user}}, \mathbf{e}_{\text{item}}, \mathbf{e}_{\text{context}}, \mathbf{e}_{\text{semantic}}, \mathbf{e}_{\text{modal}}])$ and optimizes binary cross-entropy:
\begin{equation}
\mathcal{L}_{\text{CTR}} = -\frac{1}{N} \sum_{i=1}^{N} \left[ y_i \log \hat{y}_i + (1 - y_i) \log(1 - \hat{y}_i) \right]
\end{equation}

In deployment, RQ-GMM inference runs offline in batch mode; semantic IDs are stored in feature stores for online lookup, ensuring low latency without modifying serving infrastructure.

\section{Experiments}

\subsection{Settings}

\subsubsection{Datasets and Metrics}

We use three Amazon-Review~\cite{ni2019justifying} categories: Appliances, Beauty, and Automotive. Ratings 1--3 map to 0, ratings 4--5 to 1. Data is chronologically split 8:1:1 for train/validation/test.

For the first-stage discretization task, we measure RMSE between reconstructed and original embeddings and codebook utilization (proportion of vectors used). For the second-stage CTR prediction task, we use AUC and LogLoss.

\subsubsection{Baselines and Backbones}

We compare along two dimensions: (a) discretization method—None, VQ-VAE, RQ-VAE, RQ-KMeans, RQ-GMM; (b) original modal embedding—without/with (MLP-transformed). ``None + with embedding'' directly concatenates raw BERT embeddings; ``None + without embedding'' uses only traditional features. We use FNN~\cite{zhang2016deep} and IPNN~\cite{qu2016product} as CTR backbones.

\subsubsection{Implementation Details}

All models use Adam optimizer~\cite{kingma2014adam} with batch size 1,024. Learning rate and L2 regularization are tuned in $\{10^{-6}, 10^{-5}, 10^{-4}, 10^{-3}\}$.

\textbf{First-stage discretization.} We use BERT~\cite{devlin2019bert} to generate 768-dimensional embeddings. For VQ-VAE and RQ-VAE, the encoder and decoder have hidden layers of [256, 128] with a latent dimension of 64. The VQ-VAE uses a codebook of size 256, while the RQ-VAE adopts a two-level structure with 128 codes per level. RQ-KMeans and RQ-GMM operate directly on the 768-dimensional embeddings and follow the same two-level structure (128 per level). The maximum number of iterations is set to 30, with a convergence threshold of $10^{-6}$.

\textbf{Second-stage CTR prediction.} The embedding dimension is set to 16, and the CTR prediction MLP has hidden layers of (128, 32, 8). All experiments are repeated 5 times with different random seeds, and statistical significance is assessed using paired t-tests.

\subsection{Offline Results}

\subsubsection{CTR Prediction Performance}

\begin{table}[t]
\centering
\caption{CTR prediction performance on the FNN backbone. ``w/o Emb'' and ``w/ Emb'' denote without and with the original modal embedding, respectively.}\label{tab:fnn}
\resizebox{\columnwidth}{!}{%
\begin{tabular}{ll|cc|cc|cc}
\toprule
\multirow{2}{*}{Modal Emb} & \multirow{2}{*}{Discretization} & \multicolumn{2}{c|}{Appliances} & \multicolumn{2}{c|}{Beauty} & \multicolumn{2}{c}{Automotive} \\
& & AUC & LogLoss & AUC & LogLoss & AUC & LogLoss \\
\midrule
\multirow{5}{*}{w/o Emb}
& None          & 0.643 & 0.507 & 0.594 & 0.542 & 0.613 & 0.531 \\
& VQ-VAE        & 0.651 & 0.503 & 0.599 & 0.536 & 0.620 & 0.524 \\
& RQ-VAE        & 0.657 & 0.498 & 0.607 & 0.531 & 0.627 & 0.518 \\
& RQ-KMeans     & 0.665 & 0.492 & 0.613 & 0.524 & 0.632 & 0.513 \\
& RQ-GMM        & {0.674} & {0.488} & {0.619} & {0.520} & {0.638} & {0.509} \\
\midrule
\multirow{5}{*}{w/ Emb}
& None          & 0.648 & 0.504 & 0.597 & 0.536 & 0.617 & 0.528 \\
& VQ-VAE        & 0.654 & 0.498 & 0.604 & 0.530 & 0.623 & 0.522 \\
& RQ-VAE        & 0.659 & 0.493 & 0.612 & 0.524 & 0.629 & 0.517 \\
& RQ-KMeans     & 0.667 & 0.487 & 0.621 & 0.519 & 0.635 & 0.512 \\
& RQ-GMM        & \textbf{0.678} & \textbf{0.481} & \textbf{0.628} & \textbf{0.513} & \textbf{0.644} & \textbf{0.506} \\
\bottomrule
\end{tabular}%
}
\end{table}

\begin{table}[t]
\centering
\caption{CTR prediction performance on the IPNN backbone. ``w/o Emb'' and ``w/ Emb'' denote without and with the original modal embedding, respectively.}\label{tab:ipnn}
\resizebox{\columnwidth}{!}{%
\begin{tabular}{ll|cc|cc|cc}
\toprule
\multirow{2}{*}{Modal Emb} & \multirow{2}{*}{Discretization} & \multicolumn{2}{c|}{Appliances} & \multicolumn{2}{c|}{Beauty} & \multicolumn{2}{c}{Automotive} \\
& & AUC & LogLoss & AUC & LogLoss & AUC & LogLoss \\
\midrule
\multirow{5}{*}{w/o Emb}
& None          & 0.650 & 0.498 & 0.604 & 0.533 & 0.622 & 0.522 \\
& VQ-VAE        & 0.658 & 0.493 & 0.611 & 0.526 & 0.628 & 0.515 \\
& RQ-VAE        & 0.667 & 0.490 & 0.616 & 0.524 & 0.634 & 0.509 \\
& RQ-KMeans     & 0.675 & 0.483 & 0.620 & 0.513 & 0.642 & 0.503 \\
& RQ-GMM        & {0.680} & {0.479} & {0.630} & {0.511} & {0.646} & {0.499} \\
\midrule
\multirow{5}{*}{w/ Emb}
& None          & 0.659 & 0.495 & 0.606 & 0.527 & 0.627 & 0.517 \\
& VQ-VAE        & 0.665 & 0.490 & 0.613 & 0.520 & 0.634 & 0.512 \\
& RQ-VAE        & 0.673 & 0.482 & 0.624 & 0.517 & 0.640 & 0.503 \\
& RQ-KMeans     & 0.680 & 0.481 & 0.628 & 0.511 & 0.647 & 0.496 \\
& RQ-GMM        & \textbf{0.688} & \textbf{0.472} & \textbf{0.634} & \textbf{0.505} & \textbf{0.654} & \textbf{0.494} \\
\bottomrule
\end{tabular}%
}
\end{table}

Tables \ref{tab:fnn} and \ref{tab:ipnn} present CTR prediction performance on FNN and IPNN backbones across three datasets. We observe five key findings:
(1) \textbf{Multimodal effectiveness}: Any form of multimodal information significantly outperforms no-multimodal baselines. For instance, even direct embedding concatenation yields notable AUC improvement and LogLoss reduction, validating that multimodal content captures user preferences beyond collaborative signals.
(2) \textbf{Discretization necessity}: Semantic IDs consistently improve performance both standalone and when combined with continuous embeddings, validating that discretization provides stable categorical signals naturally aligned with CTR models.
(3) \textbf{Coarse-to-fine benefit}: The ``with embedding'' variant consistently outperforms ``without embedding'' across all methods and datasets. This confirms that combining discrete semantic IDs (providing structural anchors) with continuous embeddings (preserving fine-grained nuances) helps CTR models overcome training dynamics mismatch.
(4) \textbf{RQ-GMM superiority}: RQ-GMM achieves best performance in all settings, delivering statistically significant improvements in both AUC and LogLoss over RQ-VAE and RQ-KMeans across all evaluated datasets (all $p < 0.05$). This demonstrates that higher-quality semantic IDs from probabilistic modeling translate directly to better CTR performance.
(5) \textbf{Cross-model consistency}: Results generalize across FNN and IPNN backbones, indicating fundamental advantages rather than model-specific artifacts.

\subsubsection{Discretization Quality}

\begin{table}[t]
\centering
\caption{Discretization quality comparison across datasets. Level-wise utilization rates are shown as Level 1 / Level 2.}\label{tab:quality}
\resizebox{\columnwidth}{!}{%
\begin{tabular}{l|cc|cc|cc}
\toprule
\multirow{2}{*}{Method} & \multicolumn{2}{c|}{Appliances} & \multicolumn{2}{c|}{Beauty} & \multicolumn{2}{c}{Automotive} \\
& RMSE & Util.(\%) & RMSE & Util.(\%) & RMSE & Util.(\%) \\
\midrule
VQ-VAE    & 0.614 & 33.7 & 0.621 & 23.5 & 0.587 & 41.9 \\
RQ-VAE    & 0.173 & 73.9 / 71.8 & 0.292 & 50.9 / 51.2 & 0.205 & 67.8 / 68.2 \\
RQ-KMeans & 0.121 & 86.7 / 87.1 & 0.132 & 77.3 / 78.9 & 0.119 & 83.4 / 83.1 \\
RQ-GMM    & \textbf{0.117} & \textbf{89.5 / 89.3} & \textbf{0.127} & \textbf{80.3 / 81.2} & \textbf{0.107} & \textbf{86.7 / 87.5} \\
\bottomrule
\end{tabular}%
}
\end{table}

\begin{figure}[t]
  \centering
  \includegraphics[width=\linewidth]{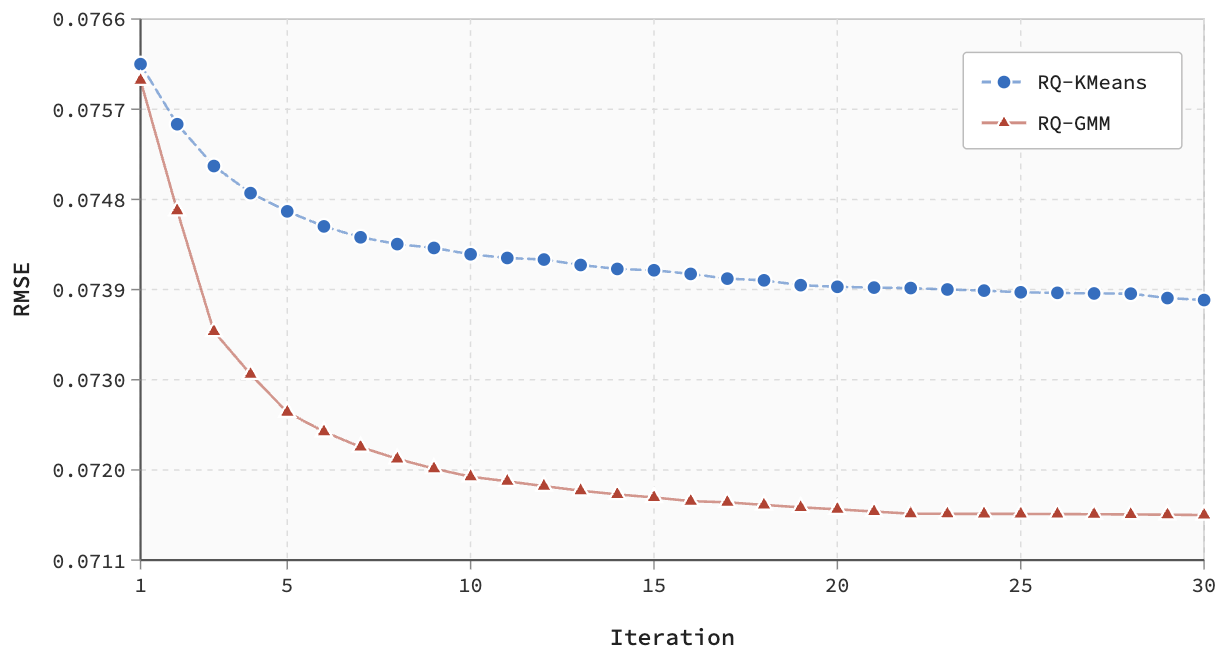}
  \caption{RMSE convergence curves of RQ-GMM and RQ-KMeans over EM/K-means iterations on a proprietary industrial dataset.}\label{fig:convergence}
  \Description{Line plots showing RMSE versus iteration number for RQ-GMM and RQ-KMeans on three datasets. RQ-GMM converges faster and to lower RMSE values with smoother curves.}
\end{figure}

Table~\ref{tab:quality} presents discretization quality metrics across three datasets. RQ-GMM consistently outperforms all baselines across all metrics. In terms of reconstruction accuracy, RQ-GMM achieves substantial RMSE reduction compared to both RQ-VAE and RQ-KMeans on average, demonstrating that probabilistic modeling better preserves semantic information during discretization.
Codebook utilization reveals striking differences. RQ-GMM achieves near-complete utilization across all datasets, substantially outperforming both RQ-VAE and RQ-KMeans. This indicates that RQ-GMM effectively avoids codebook collapse, maximizing the expressive capacity of the discrete codebook space.

Figure~\ref{fig:convergence} illustrates convergence behavior on a proprietary industrial dataset. RQ-GMM converges faster, achieves lower final RMSE, and exhibits smoother curves with less fluctuation. This reflects that the EM algorithm's soft assignment provides more stable optimization dynamics than hard K-means assignment.

\subsection{Online Results}

We conduct a 7-day A/B test on a large-scale short-video platform with hundreds of millions of daily active users, evaluating Advertiser Value (AV)—the core business metric. RQ-GMM achieves a 3.600\% improvement over direct embedding concatenation, 1.502\% over RQ-VAE, and 0.613\% over RQ-KMeans (all $p < 0.05$). These gains translate into substantial commercial value at scale. Based on these results, RQ-GMM has been fully deployed in the production system, serving daily recommendations platform-wide.

\section{Conclusions}

We propose RQ-GMM, combining Gaussian Mixture Models with residual quantization for efficient multimodal semantic discretization. Through probabilistic modeling and soft assignment, RQ-GMM better captures distributional characteristics than hard clustering methods, generating more discriminative semantic IDs. Offline and online experiments demonstrate significant improvements in reconstruction accuracy, codebook utilization, and CTR performance. RQ-GMM has been deployed at scale on a short-video platform serving hundreds of millions of daily users, providing a practical industrial solution for multimodal recommender systems. Future work includes adaptive codebook size selection across levels and unified probabilistic modeling of multiple modalities.

\begin{acks}
\end{acks}

\balance

\bibliographystyle{ACM-Reference-Format}
\bibliography{sample-base}

@article{liu2022parameter,
  title={Parameter-free dynamic graph embedding for link prediction},
  author={Liu, Jiahao and Li, Dongsheng and Gu, Hansu and Lu, Tun and Zhang, Peng and Gu, Ning},
  journal={Advances in Neural Information Processing Systems},
  volume={35},
  pages={27623--27635},
  year={2022}
}

@inproceedings{liu2023personalized,
  title={Personalized graph signal processing for collaborative filtering},
  author={Liu, Jiahao and Li, Dongsheng and Gu, Hansu and Lu, Tun and Zhang, Peng and Shang, Li and Gu, Ning},
  booktitle={Proceedings of the ACM Web Conference 2023},
  pages={1264--1272},
  year={2023}
}

@inproceedings{liu2023triple,
  title={Triple structural information modelling for accurate, explainable and interactive recommendation},
  author={Liu, Jiahao and Li, Dongsheng and Gu, Hansu and Lu, Tun and Zhang, Peng and Shang, Li and Gu, Ning},
  booktitle={Proceedings of the 46th International ACM SIGIR Conference on Research and Development in Information Retrieval},
  pages={1086--1095},
  year={2023}
}

@article{liu2023recommendation,
  title={Recommendation unlearning via matrix correction},
  author={Liu, Jiahao and Li, Dongsheng and Gu, Hansu and Lu, Tun and Wu, Jiongran and Zhang, Peng and Shang, Li and Gu, Ning},
  journal={arXiv preprint arXiv:2307.15960},
  year={2023}
}

@inproceedings{liu2023autoseqrec,
  title={Autoseqrec: Autoencoder for efficient sequential recommendation},
  author={Liu, Sijia and Liu, Jiahao and Gu, Hansu and Li, Dongsheng and Lu, Tun and Zhang, Peng and Gu, Ning},
  booktitle={Proceedings of the 32nd ACM international conference on information and knowledge management},
  pages={1493--1502},
  year={2023}
}

@inproceedings{han2025fedcia,
  title={FedCIA: Federated collaborative information aggregation for privacy-preserving recommendation},
  author={Han, Mingzhe and Li, Dongsheng and Xia, Jiafeng and Liu, Jiahao and Gu, Hansu and Zhang, Peng and Gu, Ning and Lu, Tun},
  booktitle={Proceedings of the 48th International ACM SIGIR Conference on Research and Development in Information Retrieval},
  pages={1687--1696},
  year={2025}
}

@inproceedings{liu2025filtering,
  title={Filtering Discomforting Recommendations with Large Language Models},
  author={Liu, Jiahao and Shao, Yiyang and Zhang, Peng and Li, Dongsheng and Gu, Hansu and Chen, Chao and Du, Longzhi and Lu, Tun and Gu, Ning},
  booktitle={Proceedings of the ACM on Web Conference 2025},
  pages={3639--3650},
  year={2025}
}

@article{gu2025llm,
  title={LLM-Based User Simulation for Low-Knowledge Shilling Attacks on Recommender Systems},
  author={Gu, Shengkang and Liu, Jiahao and Li, Dongsheng and Zhang, Guangping and Han, Mingzhe and Gu, Hansu and Zhang, Peng and Gu, Ning and Shang, Li and Lu, Tun},
  journal={arXiv preprint arXiv:2505.13528},
  year={2025}
}

@inproceedings{liu2025unbiased,
  title={Unbiased Collaborative Filtering with Fair Sampling},
  author={Liu, Jiahao and Li, Dongsheng and Gu, Hansu and Zhang, Peng and Lu, Tun and Shang, Li and Gu, Ning},
  booktitle={Proceedings of the 48th International ACM SIGIR Conference on Research and Development in Information Retrieval},
  pages={2555--2559},
  year={2025}
}

@inproceedings{liu2025improving,
  title={Improving LLM-powered Recommendations with Personalized Information},
  author={Liu, Jiahao and Yan, Xueshuo and Li, Dongsheng and Zhang, Guangping and Gu, Hansu and Zhang, Peng and Lu, Tun and Shang, Li and Gu, Ning},
  booktitle={Proceedings of the 48th International ACM SIGIR Conference on Research and Development in Information Retrieval},
  pages={2560--2565},
  year={2025}
}

@inproceedings{liu2025agentcf++,
  title={AgentCF++: Memory-enhanced LLM-based Agents for Popularity-aware Cross-domain Recommendations},
  author={Liu, Jiahao and Gu, Shengkang and Li, Dongsheng and Zhang, Guangping and Han, Mingzhe and Gu, Hansu and Zhang, Peng and Lu, Tun and Shang, Li and Gu, Ning},
  booktitle={Proceedings of the 48th International ACM SIGIR Conference on Research and Development in Information Retrieval},
  pages={2566--2571},
  year={2025}
}

@article{wu2025bidirectional,
  title={Bidirectional Knowledge Distillation for Enhancing Sequential Recommendation with Large Language Models},
  author={Wu, Jiongran and Liu, Jiahao and Li, Dongsheng and Zhang, Guangping and Han, Mingzhe and Gu, Hansu and Zhang, Peng and Shang, Li and Lu, Tun and Gu, Ning},
  journal={arXiv preprint arXiv:2505.18120},
  year={2025}
}

@inproceedings{zhang2025evalagent,
  title={EvalAgent: Towards Evaluating News Recommender Systems with LLM-based Agents},
  author={Zhang, Guangping and Zhang, Peng and Liu, Jiahao and Li, Zhuoheng and Li, Dongsheng and Gu, Hansu and Lu, Tun and Gu, Ning},
  booktitle={Proceedings of the 34th ACM International Conference on Information and Knowledge Management},
  pages={4086--4095},
  year={2025}
}

@article{han2026feature,
  title={Feature-Indexed Federated Recommendation with Residual-Quantized Codebooks},
  author={Han, Mingzhe and Liu, Jiahao and Li, Dongsheng and Gu, Hansu and Zhang, Peng and Gu, Ning and Lu, Tun},
  journal={arXiv preprint arXiv:2601.18570},
  year={2026}
}

@article{liu2026distribution,
  title={Distribution-Aware End-to-End Embedding for Streaming Numerical Features in Click-Through Rate Prediction},
  author={Liu, Jiahao and Ruan, Hongji and Zhang, Weimin and Tong, Ziye and Tang, Derick and Zeng, Zhanpeng and Zeng, Qinsong and Zhang, Peng and Lu, Tun and Gu, Ning},
  journal={arXiv preprint arXiv:2602.03223},
  year={2026}
}

@article{guo2017deepfm,
  title={DeepFM: a factorization-machine based neural network for CTR prediction},
  author={Guo, Huifeng and Tang, Ruiming and Ye, Yunming and Li, Zhenguo and He, Xiuqiang},
  journal={arXiv preprint arXiv:1703.04247},
  year={2017}
}

@incollection{wang2017deep,
  title={Deep \& cross network for ad click predictions},
  author={Wang, Ruoxi and Fu, Bin and Fu, Gang and Wang, Mingliang},
  booktitle={Proceedings of the ADKDD'17},
  pages={1--7},
  year={2017}
}

@inproceedings{zhou2018deep,
  title={Deep interest network for click-through rate prediction},
  author={Zhou, Guorui and Zhu, Xiaoqiang and Song, Chenru and Fan, Ying and Zhu, Han and Ma, Xiao and Yan, Yanghui and Jin, Junqi and Li, Han and Gai, Kun},
  booktitle={Proceedings of the 24th ACM SIGKDD international conference on knowledge discovery \& data mining},
  pages={1059--1068},
  year={2018}
}

@inproceedings{zhang2016deep,
  title={Deep Learning over Multi-field Categorical Data: --A Case Study on User Response Prediction},
  author={Zhang, Weinan and Du, Tianming and Wang, Jun},
  booktitle={European conference on information retrieval},
  pages={45--57},
  year={2016},
  organization={Springer}
}

@inproceedings{qu2016product,
  title={Product-based neural networks for user response prediction},
  author={Qu, Yanru and Cai, Han and Ren, Kan and Zhang, Weinan and Yu, Yong and Wen, Ying and Wang, Jun},
  booktitle={2016 IEEE 16th international conference on data mining (ICDM)},
  pages={1149--1154},
  year={2016},
  organization={IEEE}
}

@inproceedings{radford2021learning,
  title={Learning transferable visual models from natural language supervision},
  author={Radford, Alec and Kim, Jong Wook and Hallacy, Chris and Ramesh, Aditya and Goh, Gabriel and Agarwal, Sandhini and Sastry, Girish and Askell, Amanda and Mishkin, Pamela and Clark, Jack and others},
  booktitle={International conference on machine learning},
  pages={8748--8763},
  year={2021},
  organization={PmLR}
}

@inproceedings{devlin2019bert,
  title={Bert: Pre-training of deep bidirectional transformers for language understanding},
  author={Devlin, Jacob and Chang, Ming-Wei and Lee, Kenton and Toutanova, Kristina},
  booktitle={Proceedings of the 2019 conference of the North American chapter of the association for computational linguistics: human language technologies, volume 1 (long and short papers)},
  pages={4171--4186},
  year={2019}
}

@article{van2017neural,
  title={Neural discrete representation learning},
  author={Van Den Oord, Aaron and Vinyals, Oriol and others},
  journal={Advances in neural information processing systems},
  volume={30},
  year={2017}
}

@article{zeghidour2021soundstream,
  title={Soundstream: An end-to-end neural audio codec},
  author={Zeghidour, Neil and Luebs, Alejandro and Omran, Ahmed and Skoglund, Jan and Tagliasacchi, Marco},
  journal={IEEE/ACM Transactions on Audio, Speech, and Language Processing},
  volume={30},
  pages={495--507},
  year={2021},
  publisher={IEEE}
}

@inproceedings{mcqueen1967some,
  title={Some methods of classification and analysis of multivariate observations},
  author={McQueen, James B},
  booktitle={Proc. of 5th Berkeley Symposium on Math. Stat. and Prob.},
  pages={281--297},
  year={1967}
}

@article{dempster1977maximum,
  title={Maximum likelihood from incomplete data via the EM algorithm},
  author={Dempster, Arthur P and Laird, Nan M and Rubin, Donald B},
  journal={Journal of the royal statistical society: series B (methodological)},
  volume={39},
  number={1},
  pages={1--22},
  year={1977},
  publisher={Wiley Online Library}
}

@article{kingma2014adam,
  title={Adam: A method for stochastic optimization},
  author={Kingma, Diederik P},
  journal={arXiv preprint arXiv:1412.6980},
  year={2014}
}

@inproceedings{ni2019justifying,
  title={Justifying recommendations using distantly-labeled reviews and fine-grained aspects},
  author={Ni, Jianmo and Li, Jiacheng and McAuley, Julian},
  booktitle={Proceedings of the 2019 conference on empirical methods in natural language processing and the 9th international joint conference on natural language processing (EMNLP-IJCNLP)},
  pages={188--197},
  year={2019}
}

@article{deng2025onerec,
  title={Onerec: Unifying retrieve and rank with generative recommender and iterative preference alignment},
  author={Deng, Jiaxin and Wang, Shiyao and Cai, Kuo and Ren, Lejian and Hu, Qigen and Ding, Weifeng and Luo, Qiang and Zhou, Guorui},
  journal={arXiv preprint arXiv:2502.18965},
  year={2025}
}

@article{rajput2023recommender,
  title={Recommender systems with generative retrieval},
  author={Rajput, Shashank and Mehta, Nikhil and Singh, Anima and Hulikal Keshavan, Raghunandan and Vu, Trung and Heldt, Lukasz and Hong, Lichan and Tay, Yi and Tran, Vinh and Samost, Jonah and others},
  journal={Advances in Neural Information Processing Systems},
  volume={36},
  pages={10299--10315},
  year={2023}
}

@inproceedings{singh2024better,
  title={Better generalization with semantic ids: A case study in ranking for recommendations},
  author={Singh, Anima and Vu, Trung and Mehta, Nikhil and Keshavan, Raghunandan and Sathiamoorthy, Maheswaran and Zheng, Yilin and Hong, Lichan and Heldt, Lukasz and Wei, Li and Tandon, Devansh and others},
  booktitle={Proceedings of the 18th ACM Conference on Recommender Systems},
  pages={1039--1044},
  year={2024}
}

@inproceedings{sheng2024enhancing,
  title={Enhancing Taobao Display Advertising with Multimodal Representations: Challenges, Approaches and Insights},
  author={Sheng, Xiang-Rong and Yang, Feifan and Gong, Litong and Wang, Biao and Chan, Zhangming and Zhang, Yujing and Cheng, Yueyao and Zhu, Yong-Nan and Ge, Tiezheng and Zhu, Han and others},
  booktitle={Proceedings of the 33rd ACM International Conference on Information and Knowledge Management},
  pages={4858--4865},
  year={2024}
}

@inproceedings{zhang2022towards,
  title={Towards understanding the overfitting phenomenon of deep click-through rate models},
  author={Zhang, Zhao-Yu and Sheng, Xiang-Rong and Zhang, Yujing and Jiang, Biye and Han, Shuguang and Deng, Hongbo and Zheng, Bo},
  booktitle={Proceedings of the 31st ACM international conference on information \& knowledge management},
  pages={2671--2680},
  year={2022}
}

@inproceedings{luo2025qarm,
  title={Qarm: Quantitative alignment multi-modal recommendation at kuaishou},
  author={Luo, Xinchen and Cao, Jiangxia and Sun, Tianyu and Yu, Jinkai and Huang, Rui and Yuan, Wei and Lin, Hezheng and Zheng, Yichen and Wang, Shiyao and Hu, Qigen and others},
  booktitle={Proceedings of the 34th ACM International Conference on Information and Knowledge Management},
  pages={5915--5922},
  year={2025}
}

\end{document}